\documentclass[%
 reprint,
 amsmath,amssymb,
 aps,
]{revtex4-1}

\usepackage{graphicx}
\usepackage{dcolumn}
\usepackage{bm}
\usepackage{braket}

\usepackage{xcolor}

\begin{document}

\title{
Non-Hermitian skin effect in fragmented Hilbert spaces of one-dimensional fermionic lattices
}

\author{Yi-An Wang}
\affiliation{Guangdong Provincial Key Laboratory of Quantum Metrology and Sensing, School of Physics and Astronomy, Sun Yat-Sen University (Zhuhai Campus), Zhuhai 519082, China}
\author{Linhu Li}\email{lilh56@mail.sysu.edu.cn}
\affiliation{Guangdong Provincial Key Laboratory of Quantum Metrology and Sensing, School of Physics and Astronomy, Sun Yat-Sen University (Zhuhai Campus), Zhuhai 519082, China}

\date{\today}

\begin{abstract}
We discover that the interplay between Hilbert space fragmentation and multiple non-Hermitian pumping channels leads to distinct non-Hermitian skin effect (NHSE) in real and Fock spaces. Using an extended Hatano-Nelson model with next-nearest neighbor hopping and a strong interaction as an example, we find that two fermions loaded in the lattice exhibit different real-space NHSE depending on the Hilbert space fragments they belong to. Moreover, in the high-energy sector resulting from the fragmentation, the two-particle bound states form a one-dimensional lattice in Fock space, resulting in the Fock-space NHSE. At half-filling, while real-space NHSE is suppressed by many-body effects, richer patterns of Fock-space skin-like localization are found to emerge for different fragmented energy sectors and subsectors. This work extends our understanding of the interplay between NHSE and Hilbert space fragmentation and provides detailed insights into their manifestation in interacting non-Hermitian systems.
\end{abstract}

\maketitle


\section{\label{sec:level1}Introduction}
Within the framework of non-Hermitian physics \cite{Ashida:2020dkc}, many novel phenomena emerge that have no counterpart in Hermitian systems \cite{PhysRevLett.121.086803, PhysRevLett.121.026808, PhysRevLett.121.136802, PhysRevLett.123.066404, PhysRevLett.122.076801, PhysRevLett.123.206404, PhysRevLett.123.066405, PhysRevB.100.054105, minganti2019quantum, PhysRevLett.123.016805, PhysRevX.9.041015, PhysRevLett.125.126402, li2020critical, PhysRevLett.124.056802, PhysRevLett.124.086801, PhysRevLett.124.250402, RevModPhys.93.015005, zhang2021acoustic,PhysRevLett.128.223903, zhang2022universal, li2022non, PhysRevB.107.L220301, PhysRevX.13.021007, PhysRevLett.131.036402, lin2023topological, li2023enhancement, okuma2023non, qin2024geometry, li2024localization, yuce2024strong, lin2024topologically}. One of the most striking phenomena is the non-Hermitian skin effect (NHSE), where all eigenstates are localized non-reciprocally at the boundaries of the system \cite{PhysRevLett.121.086803, PhysRevLett.121.026808, PhysRevLett.121.136802, PhysRevLett.123.066404, PhysRevLett.122.076801, PhysRevLett.123.206404, PhysRevLett.123.066405}.
Based on this foundation, even at the single-particle level, numerous other intriguing phenomena have been found to arise from the interplay between multiple skin channels of different components or hopping ranges,
such as critical NHSE \cite{li2020critical}, bipolar NHSE \cite{zhang2021acoustic}, and the non-Hermitian pseudo-gap \cite{li2022non}.

The scope of NHSE has been greatly extended by including many-body effects \cite{PhysRevLett.123.123601, PhysRevB.102.235151, lee2021many, shen2022non, faugno2022interaction, li2023many, longhi2023spectral, mao2023non, qin2024occupation, yoshida2024non, brighi2024nonreciprocal, qin2024dynamical, kim2024collective, Shen:2024doc, PhysRevLett.132.086502,PhysRevLett.133.136502,PhysRevLett.133.136503,li2024dissipation},
resulting in more exotic variations, such as skin localization acting on the spin degree of freedom \cite{hamanaka2024multifractality} or multiparticle Fock space \cite{PhysRevLett.133.136502, Shen:2024doc}. On the other hand, a strong many-body interaction may break the Hilbert space into an exponentially large number of subsectors with respect to the system size, known as the Hilbert Space Fragmentation (HSF) \cite{PhysRevLett.124.207602, moudgalya2022quantum, francica2023hilbert, ghosh2024hilbert, kumar2024hilbert, de2019dynamics, frey2022hilbert, herviou2021many}.
In non-Hermitian systems, HSF has been found to impose a real spectrum even without $PT$ symmetry \cite{ghosh2024hilbert},
and is closely related to rich classes of skin clusters with distinct localization patterns \cite{shen2022non}. 
A natural question arises: In non-Hermitian systems exhibiting HSF, what are the characteristics of NHSE in both real space and Fock space for different subsectors?

In this work, we consider a non-Hermitian one-dimensional (1D) lattice with multiple skin channels induced by nearest- and next-nearest-neighbor non-reciprocal hopping terms, and a strong nearest-neighbor (NN) interaction that facilitates HSF. We first discuss the minimal case with $N=2$ particles, where the Hilbert space is fragmented into two sectors separated from each other in real energies. Particles in the two sectors are influenced by different non-reciprocal hopping terms and exhibit different NHSE,
manifesting the occupation-dependent particle separation \cite{qin2024occupation}. By treating each Fock basis as an effective lattice site, the high-energy sector can be mapped to a 1D non-Hermitian lattice, exhibiting the Fock skin effect. We then extend our study to the half-filled case, where energy sectors are further fragmented into several disconnected subsectors with similar eigenenergies. We find that, although real-space NHSE is suppressed by the fermionic statistic, skin-like localization still emerges in Fock space for each subsector.

The rest of this paper is organized as follows. In Sec. \ref{Sec:model} we introduce the one-dimensional model Hamiltonian and reveal the NHSE at the single-particle level of the system. Sec. \ref{Sec:two-particle} and Sec. \ref{Sec:half-filling} are devoted to the two-particle and half-filling scenarios, respectively,  where skin localization in both real space and Fock space is studied in detail. A summary of our results is given in Sec. \ref{Sec:summary}.

\section{Model and single-particle NHSE}\label{Sec:model}
We consider a one-dimensional non-Hermitian fermionic lattice system described by the Hamiltonian:
\begin{figure}
    \centering
    \includegraphics[width=1.0\linewidth]{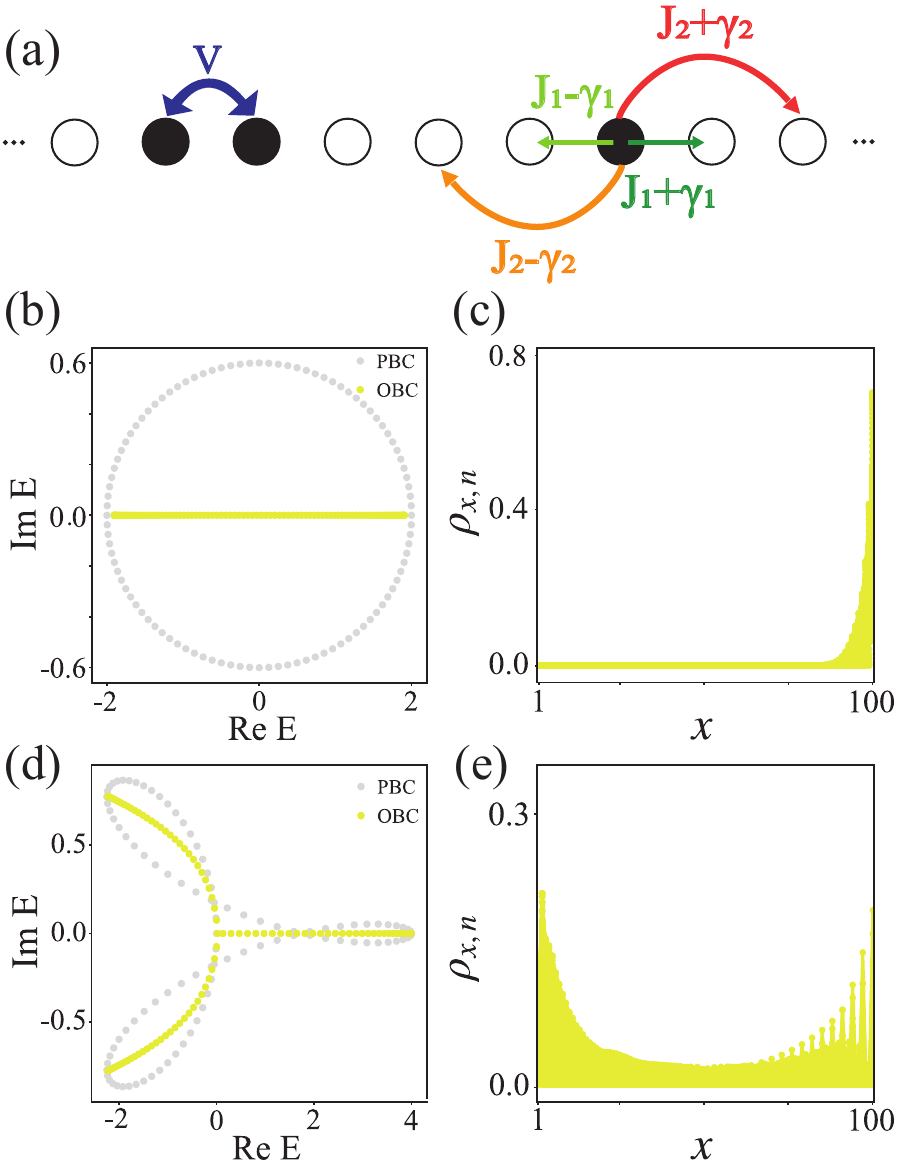}
    \caption{
    (a) A schematic diagram of the extended HN model. $J_1 \pm \gamma_1 $ and $J_2 \pm \gamma_2 $ are the hopping amplitudes for NN and NNN hopping, respectively, and $V$ measures the interaction strength.
    (b) shows the single-particle energy spectra under PBCs (gray dots) and OBCs (yellow dots), 
    considering only NN hopping. 
    (c) depicts the eigenstate distributions under OBCs.
    (d) and (e) present the same information as (b) and (c), except with both NN and NNN hopping.
    In both cases, nontrivial spectral winding is observed, indicating the presence of NHSE. 
    Parameters are  (b) $J_1 = 1.0, \gamma_1 = 0.9, J_2 = \gamma_2 = 0$, and (d) $J_1 = J_2 = 1.0, \gamma_1 = 0.9, \gamma_2 = -0.6$.}
    \label{fig:model}
\end{figure}
\begin{align}
    \hat{H} &= \sum_{x=1}^{L-1} [ (J_1 - \gamma_1)\hat{c}_x^{\dagger}\hat{c}_{x+1} + (J_1 + \gamma_1)\hat{c}_{x+1}^{\dagger}\hat{c}_{x}] \cr
    &+ \sum_{x=1}^{L-2} [(J_2 - \gamma_2)\hat{c}_x^{\dagger}\hat{c}_{x+2} + (J_2 + \gamma_2)\hat{c}_{x+2}^{\dagger}\hat{c}_{x}] \cr
   &+ V\sum_{x=1}^{L-1}\hat{n}_{x}\hat{n}_{x+1},\label{eq:H}
\end{align}
namely an extended Hatano-Nelson (HN) model \cite{PhysRevLett.77.570} with both nearest- and next-nearest-neighbor (NN and NNN) hoppings and a nearest-neighbor interaction, as shown in Fig.~\ref{fig:model}(a). 
The operators $\hat{c}_x$ and $\hat{c}_x^{\dagger}$ represent the fermionic annihilation and creation operators at the lattice site $x$, respectively, and $\hat{n}_x = \hat{c}_x^{\dagger}\hat{c}_x$ denotes the particle number operator at that site. 
$J_{1,2}$ and $\gamma_{1,2}$ represent the amplitudes of NN and NNN hopping with $\gamma_{1,2}$ the non-reciprocal terms,
$V$ measures the strength of the interaction term, and $L$ is the length of the lattice. At the single-particle level,  the NN and NNN hopping can be viewed as two non-reciprocal ``channels" for the particle to travel along, which lead to NHSE with different localization profiles and topological features in different parameter regimes. Specifically, when $J_2=\gamma_2=0$, NN hopping alone induces a nontrivial spectral winding topology under the periodic boundary conditions (PBCs), with unidirectional NHSE under the open boundary conditions (OBCs), as shown in Fig. \ref{fig:model}(b) and (c). The presence of non-reciprocal NNN hopping may further induce a bipolar NHSE~\cite{zhang2021acoustic} 
with self-crossing spectral trajectory under PBCs, provided that the NN and NNN hopping have stronger magnitudes toward opposite directions,
as shown in Fig. \ref{fig:model}(d) and (e).

In the many-body scenario with a strong interaction $V$, we can perform a perturbative expansion of the Hamiltonian, retaining terms up to the first order in $(J_1, \gamma_1, J_2, \gamma_2)$. An effective projected Hamiltonian can be obtained as \cite{PhysRevB.62.7791}
\begin{align}
    \hat{H} &\rightarrow \hat{H}_{\rm eff} \cr
    &= \sum_{x=1} \hat{P}_x^{(1)}\left( (J_1 - \gamma_1)\hat{c}_x^{\dagger}\hat{c}_{x+1} + (J_1 + \gamma_1)\hat{c}_{x+1}^{\dagger}\hat{c}_{x}\right) \hat{P}_x^{(1)}\cr
    &+ \sum_{x=1} \hat{P}_x^{(2)}\left( (J_2 - \gamma_2)\hat{c}_x^{\dagger}\hat{c}_{x+2} + (J_2 + \gamma_2)\hat{c}_{x+2}^{\dagger}\hat{c}_{x}\right) \hat{P}_x^{(2)}\cr
    &+ V\sum_{x=1}^{L-1}\hat{n}_{x}\hat{n}_{x+1},\label{eq:Heff}
\end{align}
where $\hat{P}_x^{(1)} = 1 - (\hat{n}_{x-1} - \hat{n}_{x+2})^2$ and
$\hat{P}_x^{(2)} = 1 - (\hat{n}_{x-1} - \hat{n}_{x+3})^2$ are two projection operators. 
Namely, the strong interaction imposes significant constraints on the particle's hopping, 
such that NN hopping can only occur when $\left<\hat{n}_{x-1}\right> = \left<\hat{n}_{x+2}\right>$ and NNN hopping can only occur when $\left<\hat{n}_{x-1}\right> = \left<\hat{n}_{x+3}\right>$. 
Thus the complete Hilbert space is fragmented into energetically separated sectors, each formed by Fock states with the same number of pairs of adjacent particles ($\sum_x \hat{n}_{x}\hat{n}_{x+1}$), manifesting the HSF.

\section{Occupation-dependent NHSE for two particles}\label{Sec:two-particle}
\subsection{Real-space NHSE}
\begin{figure*}
    \centering
    \includegraphics[width=1.0\linewidth]{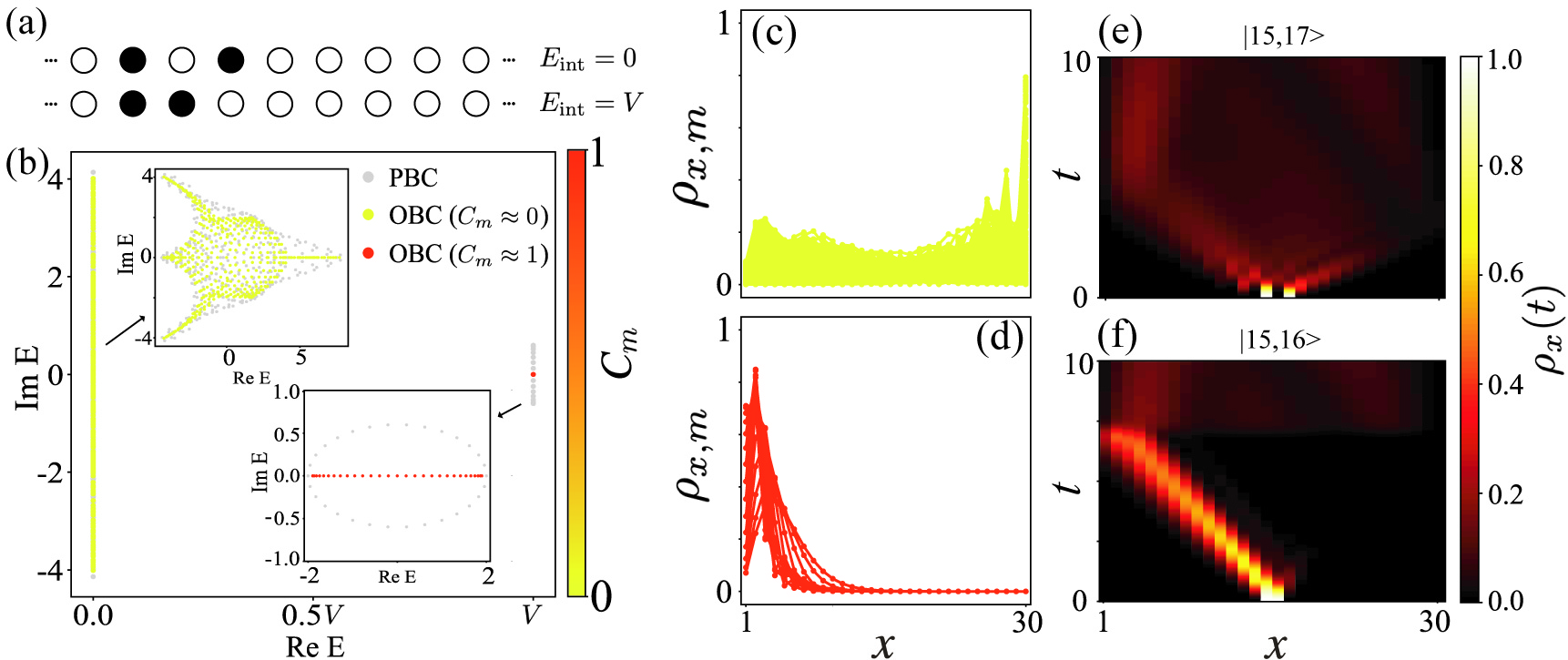}
   \caption{
    (a) Occupation configurations of $N=2$ particles with interaction energy $E_{\rm int}=0$ and $V$, respectively.
    (b) Energy spectrum for $N=2$ under PBCs (gray dots) and OBCs (colored). Colors map the value of the nearest-neighbor correlation $C_m$ for each eigenenergy. The eigenenergies are fragmented into two sectors along the real part of the energy. The insets show detailed views of the low-energy and high-energy sectors. 
    (c) and (d) show the particle density distributions in the low-energy and high-energy sectors [with the same colors in (b)], respectively. In the low-energy sector, the density distribution shows a bipolar distribution, with stronger localization to the right.
    In the high-energy sector, the distribution is localized unidirectionally to the left. 
    (e) and (f) display the time-evolution for initial states with two particles arranged at sites $(15,17)$ and $(15,16)$, respectively.
    Two separated particles are seen to diffuse in the lattice, and the adjacent ones are pumped to the left.
    Parameters are $L = 30, J_1 = J_2 = 1.0, \gamma_1 = 0.9, \gamma_2 = -0.3, V = 10^{10}$ in all panels.}
    \label{fig:N2_eigen}
\end{figure*}

We first discuss the minimal case with $N = 2$ particles, where Fock states contain either one or none adjacent pair of particles with interaction energy $E_{\rm int} =\langle V\sum_{x=1}^{L-1}\hat{n}_{x}\hat{n}_{x+1}\rangle=V$ or $0$, respectively, as sketched in Fig. \ref{fig:N2_eigen}(a).
Thus, a strong interaction $V$ is expected to fragment the Hilbert space into two subspaces accordingly.
By numerically solving the eigenvalue problem of the Hamiltonian $\hat{H}|\psi_m\rangle=E_m|\psi_m\rangle$ with $m$ indexing different eigenstates, 
we obtain the energy spectrum and the particle density distribution 
$$\rho_{x,m} = \bra{\psi_m} \hat{n}_x \ket{\psi_m}$$ 
of the system, 
as shown in Fig.~\ref{fig:N2_eigen}(b) to (d). 
The eigenenergies in Fig. \ref{fig:N2_eigen}(b) are seen to be separated into two sectors in terms of their real parts.
Note that a very strong interaction ($V=10^{10}$) is considered in our numerics, which ensures the HSF and that the NN hopping does not affect the subspace with two or more adjacent particles, as discussed in Appendix \ref{app:N2}. 
To verify the HSF of the energy sectors, we define a nearest-neighbor correlation function
\begin{align}
    C_m = \sum_{x=1}^{L-1}\bra{\psi_m}\hat{n}_x\hat{n}_{x+1}\ket{\psi_m},
\end{align}
which ``counts" the number of adjacent pair of particles in $|\psi_m\rangle$.
It is seen in Fig. \ref{fig:N2_eigen}(b) that $C_m\approx 0$ and $1$ for the low-energy and high-energy sectors, respectively, confirming a nearly perfect HSF of our system.

In addition to separating eigenenergies into different sectors,  the HSF further restricts the effective non-reciprocal hopping for the Fock states, leading to qualitatively different skin localization for the two sectors, as shown in Fig.~\ref{fig:N2_eigen}(c) and (d).
Explicitly, the high-energy sector corresponds to particles occupying two adjacent lattice sites.
Therefore it  is evidently influenced only by the NNN hopping, as NN hopping will inevitably separate the two adjacent particles and couple the two fragmented Hilbert space with different real energies. 
This can be further elaborated by performing first-order perturbation theory, projecting the Hamiltonian onto the high-energy subspace, resulting in an effective Hamiltonian
\begin{align}
     &\hat{H}_{\rm eff} =\hat{P}_{\rm int} \hat{H}\hat{P}_{\rm int}\cr   
      &=V+\hat{P}_{\rm int} \left(\sum_{x=1}^{L-2} [(J_2 - \gamma_2)\hat{c}_x^{\dagger}\hat{c}_{x+2} + (J_2 + \gamma_2)\hat{c}_{x+2}^{\dagger}\hat{c}_{x}] \right)\hat{P}_{\rm int }\cr
\end{align}
where $\hat{P}_{\rm int} = \sum_{x=1}^{L-1} \ket{\alpha_x}\bra{\alpha_x}$ with $\ket{\alpha_x} =\hat{c}_x^\dagger\hat{c}_{x+1}^{\dagger}\ket{0}$ projects the Hamiltonian to the subspace with adjacent pairs of particles
(see Appendix \ref{app:N2} for detailed derivation).
Thus, skin localization toward the left (right) emerges when $|J_2-\gamma_2|>|J_2+\gamma_2|$ ($|J_2-\gamma_2|<|J_2+\gamma_2|$), as shown in Fig. \ref{fig:N2_eigen}(d).
In contrast, eigenstates in the low-energy sector live in the subspace with particles occupying non-adjacent sites, and are subjected to both NN and NNN hopping. 
As in the single-particle picture, low-energy eigenstates generally display a bipolar localization when the non-reciprocity of NN and NNN takes opposite direction. 
Further requiring a stronger non-reciprocal strength for the NN hopping, these states with separated particles can be made mostly localized at the opposite direction to the adjacent particles [Fig. \ref{fig:N2_eigen}(c)].
\begin{figure}
    \centering
    \includegraphics[width=1.0\linewidth]{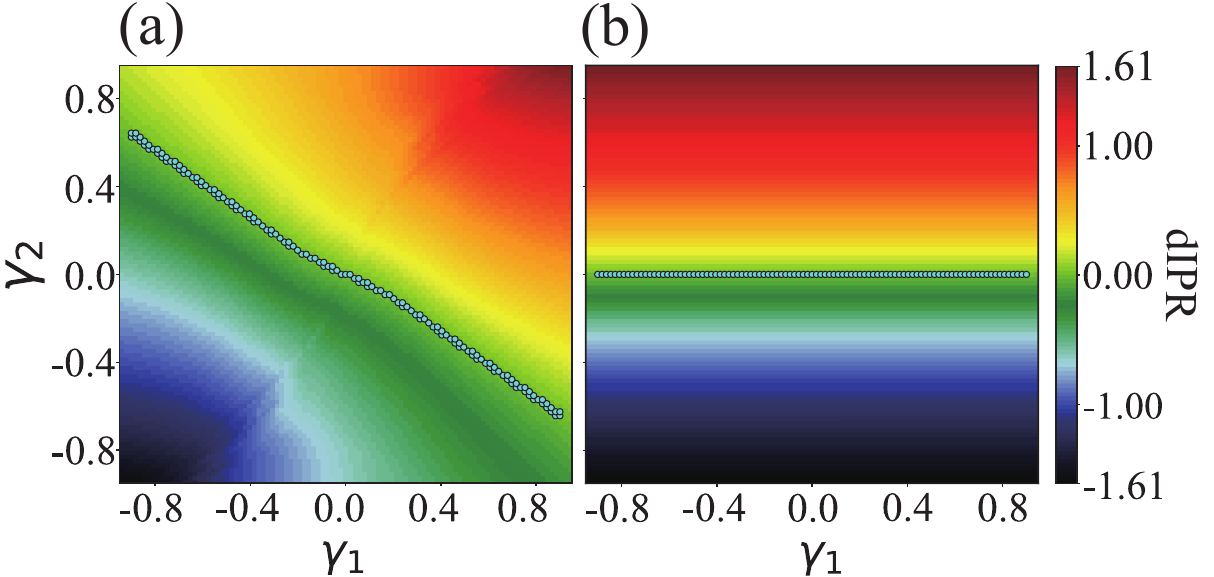}
    \caption{(a) and (b) show the ${\rm dIPR}$ for the low-energy and high-energy sectors, respectively, with different dependence on $\gamma_1$ and $\gamma_2$.
    Scatter points in the figures highlight the regions with ${\rm dIPR} = 0$. 
    Localization to the left (right) corresponds to negative (positive) ${\rm dIPR}$, and the degree of localization is reflected by its magnitude.
    Parameters (except $\gamma_{1,2}$) are the same as in Fig. \ref{fig:N2_eigen}.
    }
    \label{fig:N2_diagram}
\end{figure}

In Fig. \ref{fig:N2_eigen}(e) and (f), we display the dynamics of different paired and unpaired states as a manifestation of the occupation-dependent real-space NHSE. 
The initial state is chosen as $|\psi(0)\rangle=|x_1,x_2\rangle=\hat{c}^\dagger_{x_1}\hat{c}^\dagger_{x_2}|0\rangle$ with the two particles placed at sites $x_1$ and $x_2$, where $x_2>x_1$ is set without loss of generality.
The normalized distribution of the state, after evolving under the Hamiltonian $\hat{H}$ for a period of time $t$,
is given by 
\begin{eqnarray}
\rho_x(t)=\langle\psi(t)|\hat{n}_x |\psi(t)\rangle
\end{eqnarray}
with
\begin{align}
    \ket{\psi(t)} = \frac{e^{\frac{-i\hat{H}t}{\hbar}}\ket{\psi(0)}}{\sqrt{\bra{\psi(0)}e^{\frac{i\hat{H}^{\dagger}t}{\hbar}}e^{\frac{-i\hat{H}t}{\hbar}}\ket{\psi(0)}}}.
\end{align}
Due to the competition between NN and NNN hoppings, the unpaired initial state ($x_2>x_1+1$) does not show a clear left- or right-moving tendency during the evolution, but diffuses throughout the lattice over time. On the other hand, two adjacent particles ($x_2=x_1+1$) show a unidirectional ballistic evolution toward the left, in consistent with the 
skin localization for static eigenstates in the high-energy sector.

To characterize the skin localization for the two sectors,
we define the average directional inverse participation ratio (dIPR) for many-body eigenstates
\begin{align}
{\rm dIPR} = \frac{1}{N_s}\sum_{n=1}^{N_s}\sum_{x = 1}^{L}\frac{x-\frac{L+1}{2}}{\frac{L-1}{2}}\rho_{x,n}^2,
\end{align}
which reflects both the degree and direction of localization in the concerned sector with $N_s$ eigenstates.
${\rm dIPR} > 0$ indicates that the distribution of particle number density is mainly on the right side of the system, while ${\rm dIPR} < 0$ indicates that it is primarily on the left side. In both cases, a larger magnitude of dIPR indicates a stronger edge localization toward the corresponding direction.
As can be seen in Fig. \ref{fig:N2_diagram},
the average dIPR of the low-energy sector varies with both $\gamma_1$ and $\gamma_2$, and a strong localization occurs when the two non-reciprocal parameters have the same sign.
In the high-energy sector, however, the degree and direction of localization are determined solely by the NNN hopping parameter $\gamma_2$. These numerical results further validate the different real-space NHSE induced by HSF in our model.

\subsection{Fock-space NHSE}
\begin{figure}
    \centering
    \includegraphics[width=1.0\linewidth]{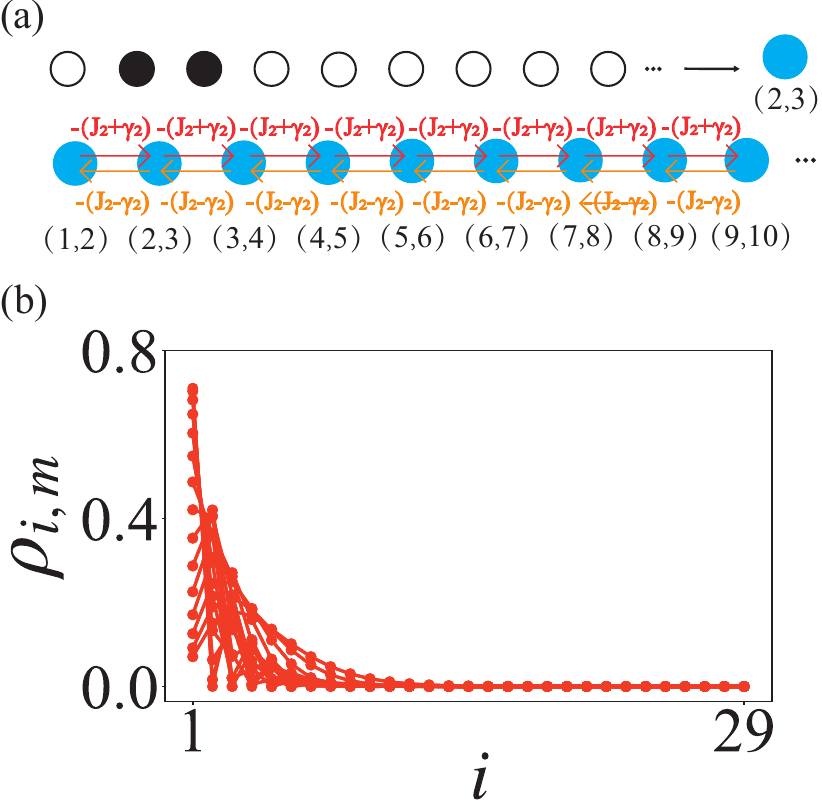}
    \caption{
    (a) Illustration of the abstract Fock lattice in the subspace of two adjacent particles. $(x_1,x_2)$ indicate the Fock state with two particles occupying sites $x_1$ and $x_2$, respectively.
    Hopping amplitudes between abstract lattice sites are $-(J_2 + \gamma_2)$ to the right and $-(J_2 - \gamma_2)$ to the left.
    (b) Distribution in the Fock space for eigenstates in the high-energy sector.
     Parameters are the same as in Fig. \ref{fig:N2_eigen}.}\label{fig:N2_Fock}
\end{figure}

Subjected only to the NNN hopping, the high-energy sector can be mapped onto an abstract 1D lattice, with each lattice site corresponding to a Fock state with a pair of adjacent particles, as sketched in Fig. \ref{fig:N2_Fock}(a).
It is essentially the non-Hermitian HN model with non-reciprocal hopping given by $J_2\pm \gamma_2$, where skin localization emerges with nonzero $J_2$ and $\gamma_2$.
In Fig. \ref{fig:N2_Fock}(b), we demonstrate the distribution in the abstract lattice of eigenstates of the high-energy sector, 
\begin{eqnarray}
\rho_{i,m} = |\langle\alpha_i|\psi_m\rangle|^2
\end{eqnarray}
with $i$ labeling different two-particle Fock state. 
Skin localization toward the left can be clearly seen when $\gamma_2<0$, indicating the Fock-space NHSE that drives the particles to $|\alpha_1\rangle=\hat{c}^\dagger_1\hat{c}^\dagger_2|0\rangle$.

\section{Occupation-dependent NHSE at half-filling}\label{Sec:half-filling}
\begin{figure*}
    \centering
    \includegraphics[width=1.0\linewidth]{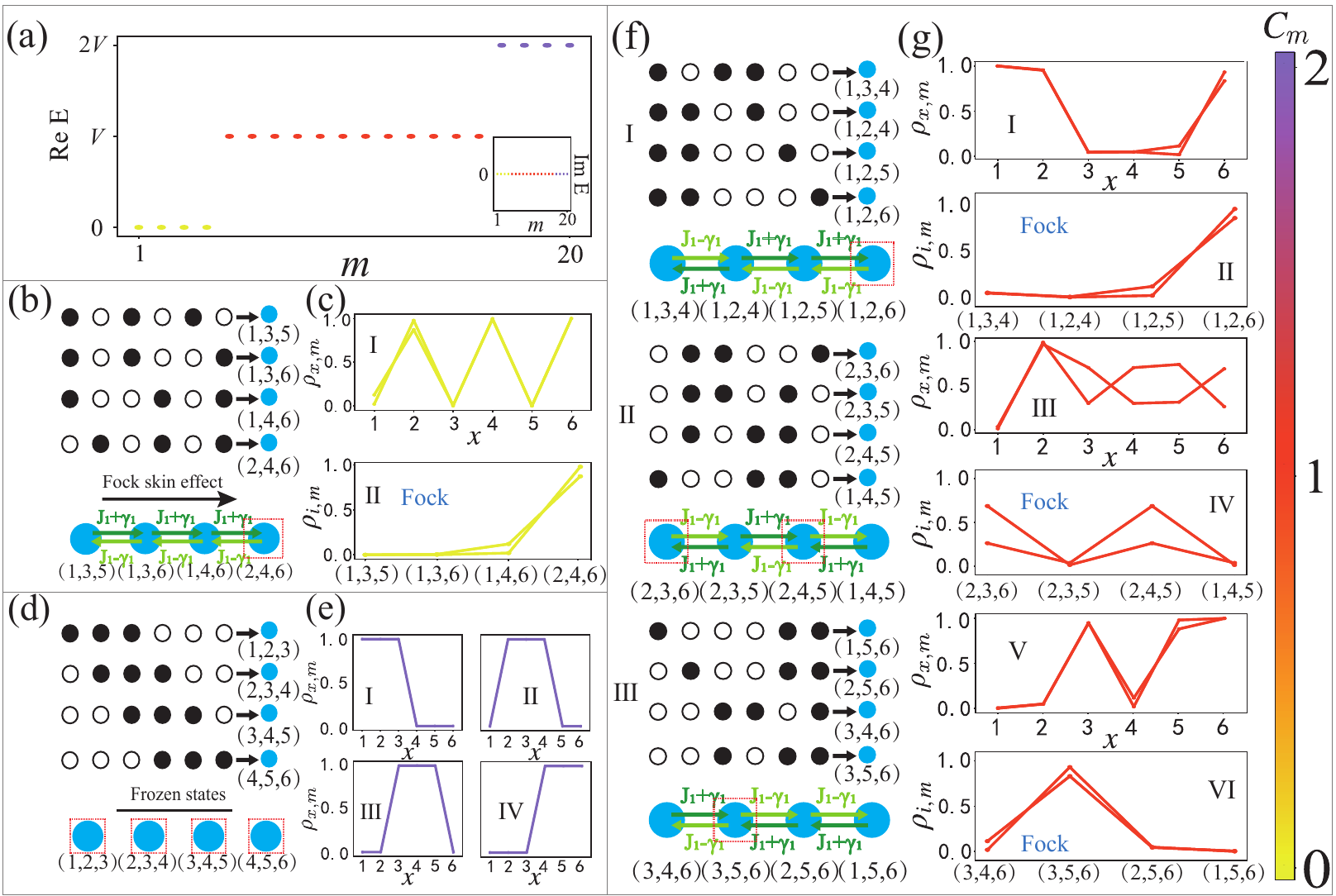}
    \caption{
    (a) Real parts of eigenenergies for the half-filled lattice with $L=6$, divided into three sectors. Colors indicate the nearest-neighbor correlation for each eigenstate (as shown on the right-hand side of the figure). Inset displays the imaginary parts of eigenenergies, which are all zero.
    (b) The abstract lattice in Fock space for the lowest-energy sector. 
    Red dashed box indicates the abstract lattice site where the eigenstates are localized.
    (c) The particle density distribution of the lowest-energy sector in real space and Fock space, respectively.
    (d) and (e) show similar results as in (b) and (c), but for the highest-energy sector. Only real-space distribution of eigenstates are shown in (e), as the four abstract lattice sites in the Fock space are decoupled from each other in this subspace.
    (f) and (g) show similar results as in (b) and (c), but for the intermediate-energy sector.
    The intermediate-energy sector is further fragmented into three decoupled subspaces under a strong $V$.
    Parameters are $L = 30, J_1 =  1.0, \gamma_1 = 0.9, J_2=\gamma_2 = 0, V = 10^{10}$.
    }\label{fig:half_NN}
\end{figure*}
With more particles loaded in the lattice,
the numbers of fragmented subspaces and corresponding energy sectors also increase subsequently, 
greatly enriching the particle distribution profiles of different sectors compared to the two-particle case.
To give a concrete example,
here we focus on the  system at half-filling with the particle number $N = \frac{L}{2}$, and $L=6$ lattice sites to demonstrate the most typical features of NHSE therein.

\subsection{With only NN hopping}
We first consider the case with only NN hopping ($J_2=\gamma_2=0$).
 According to~Eq. \eqref{eq:Heff}, a strong interaction $V$ constrains allows NN hopping to occur only between Fock states with $\left<\hat{n}_{x-1}\right> = \left<\hat{n}_{x+2}\right>$, dividing the system into three energy sectors, as shown in Fig. \ref{fig:half_NN}(a).
Among them, the HSF is most clear for the lowest- and highest-energy sectors ($E\approx 0$ and $2V$).
The lowest-energy sector is spanned by four Fock states with no adjacent particles, which correspond to the four occupation configurations in~Fig. \ref{fig:half_NN}(b).
Their abstract lattice forms the HN model with non-reciprocal hopping given by $J_1\pm \gamma_1$, 
exhibiting the Fock-space NHSE.
Explicitly, as demonstrated in Fig. \ref{fig:half_NN}(c) with $|J_1+\gamma_1|>|J_1-\gamma_1|$, eigenstates in the lowest-energy sector localize toward the Fock state labeled by $(2,4,6)$ (the real-space positions of particles), showing a density-wave distribution in real space and a skin localization in Fock space.
On the other hand,
the subspace of the highest-energy sector also contains four Fock states as its basis, 
which are decoupled from each other due to the constraint of Eq. \eqref{eq:Heff}, as shown in Fig. \ref{fig:half_NN}(d).
Thus, eigenstates in this sector are ``frozen"  at each of them, forming disconnected subsectors disconnected with particles uniformly distributed on $N$ adjacent lattice sites in real space [Fig. \ref{fig:half_NN}(d)].

Next, we consider the case in the intermediate-energy sector. 
Since NN hopping can only occur when $\left<\hat{n}_{x-1}\right> = \left< \hat{n}_{x+2}\right>$ under a strong interaction $V$,
this sector is further fragmented into three degenerate subsectors disconnected from each other, each spanned by four Fock states as its basis, as illustrated in Fig.~\ref{fig:half_NN}(f).
Unlike the lowest-energy sector, 
the abstract Fock lattices of the intermediate-energy sector do not possess unidirectional non-reciprocity, resulting in different distribution patterns of particles.
Numerically, particles are found to mostly occupy the Fock state with the longest hopping sequence possessing unidirectional non-reciprocity in the abstract lattice,
as shown in Fig.~\ref{fig:half_NN}(g).
For example, the abstract lattice in Fig.~\ref{fig:half_NN}(f-I) possesses two consecutive stronger hopping ($J_1+\gamma_1$) toward the Fock state $(1,2,6)$, giving raise to the skin-like localization at this state in Fig.~\ref{fig:half_NN}(g-I,II).

\subsection{With both NN and NNN hopping}
\begin{figure}
    \centering
    \includegraphics[width=1.0\linewidth]{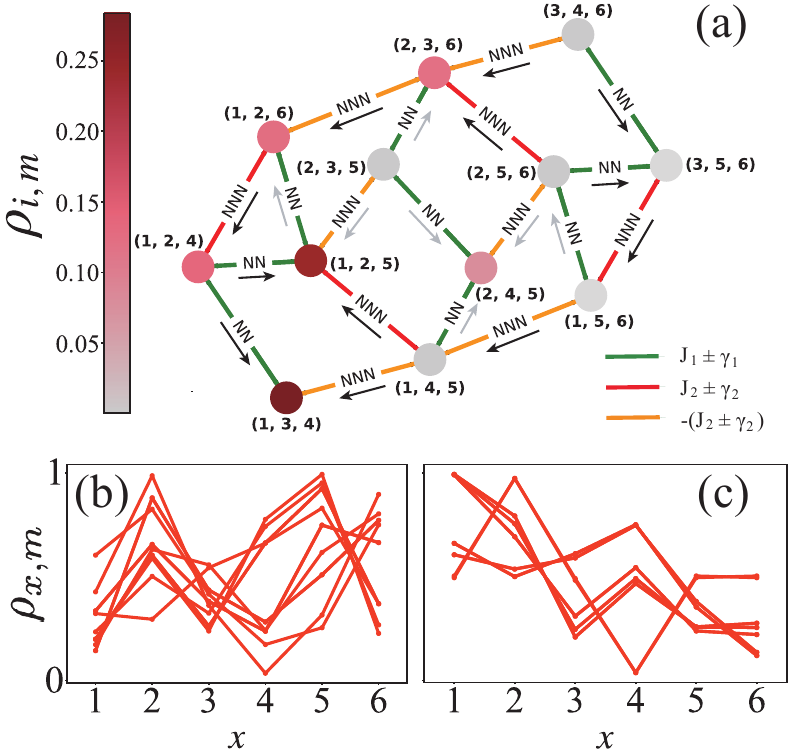}
    \caption{(a) Connectivity graph of the abstract lattice in Fock space for the intermediate-energy sector with $L=6$ at half-filling, in the presence of both NN and NNN hopping.
    Colors mark the hopping amplitudes between different abstract lattice sites. 
    Arrows (both black and gray) indicate the direction of stronger hopping amplitudes for the parameters we choose. Black arrows highlight the longest sequences (four sites) of hopping with unidirectional non-reciprocity.
     (b) and (c) show the particle density distributions in real space for $\gamma_2=0$ and $\gamma_2=-0.9$, respectively, and $J_2=1$.
    Other parameters are the same as in Fig. \ref{fig:half_NN}.
    }\label{fig:half_NNN}
\end{figure}

At half-filling, NNN non-reciprocal hopping affects the eigenstate distribution rather differently compared to the two-particle case.
Due to the HSF, NNN hopping has no effect on the highest- and lowest-energy sectors, leading to the same subspaces and localization of eigenstates as in Fig. \ref{fig:half_NN}(b) to (e). 
However, the three subsectors in Fig. \ref{fig:half_NN}(f) are jointed together by NNN hopping, 
resulting in a more complex abstract lattice model without a well-defined crystal structure, as  shown in Fig.~\ref{fig:half_NNN}(a).
Nevertheless, skin-like localization can still be observed in the Fock space,
toward the abstract lattice sites with the longest hopping sequences with unidirectional non-reciprocity.
Specifically in our example with $L=6$, 
such hopping sequences start from the Fock states $(2,5,6)$ and $(3,5,6)$ and end at $(1,2,5)$ and $(1,3,4)$, resulting in the localization of many-body eigenstates at the latter.
Here, we set $\gamma_1=-\gamma_2$ to ensure that the non-reciprocal strengths for NN and NNN hopping are equal. 
If this condition were not met, the non-reciprocity of certain hopping could dominate, leading to distinct skin-like localization patterns in Fock space primarily driven by the hopping with the greater non-reciprocal strength (not shown).

In real space, NHSE is significantly suppressed by fermionic statistics, which can already be seen in Fig. \ref{fig:half_NN} with only NN hopping.
This suppression occurs because a pair of adjacent fermions becomes nearly frozen in the lattice due to the HSF; they can only move when there is another particle separated from the pair by one lattice site.  
For instance, in Fig. \ref{fig:half_NN}(f-I), the adjacent pair at sites $(1,2)$ can only shift if there is another particle at site $4$, allowing a particle to hop from site $2$ to $3$, thus forming a new pair at $(3,4)$. 
Nevertheless, longer-range hopping enhances the mobility of the particles, as it requires more of them to be adjacent to form such a nearly-frozen set of particles,  which is also easier to unfreeze since longer-range hopping allows the particles to hop across multiple lattice sites to interact with others.
Indeed, as seen in Fig. \ref{fig:half_NNN}(b) and (c), 
non-reciprocal NN hopping alone does not induce a clear skin-localization in real space, even in the presence of reciprocal (Hermitian) NNN hopping;
and particles are seen to accumulate to an edge of the lattice when further introducing non-reciprocity to NNN hopping.

Finally, we note that here we have only demonstrated results for $L=6$ lattice sites, yet they already unveil the most typical features of skin localization of the system at half-filling.
For larger systems, the fragmented subspaces for 
the highest- and lowest-energy sectors are spanned by Fock states similar to that of Fig. \ref{fig:half_NN}(b) and (d),
thus exhibiting the same properties of Fock-space NHSE and frozen states, respectively, as shown in Appendix \ref{sec:large_L}.
On the other hand, more intermediate sectors with different real energies emerge, as the number of adjacent pairs of particles ranges from $0$ to $L/2-1$.
Similar to the case of $L=6$,
the distribution of eigenstates is highly dependent on the occupation configurations of particles in each sector. As the system size increases, these configurations become significantly more diverse, necessitating a case-by-case analysis to fully understand their implications.

\section{Summary}\label{Sec:summary}
In this work, we investigate the emergence of HSF induced by a strong nearest-neighbor interaction $V$ in a one-dimensional fermionic lattice with NN and NNN hopping,
revealing intriguing interplays between HSF and multiple non-Hermitian pumping channels for two scenarios with different numbers of particles. 

In the case with $N=2$ particles, the energy spectrum is fragmented into two distinct sectors. Restricted by particle configurations allowed in each sector, particle density distribution in the low-energy sector is influenced by both NN and NNN non-reciprocal hopping terms, and the high-energy sector is solely affected by NNN hopping, showcasing distinct real-space NHSE. 
In addition, the high-energy sector can be mapped to a one-dimensional HN model based on Fock basis, resulting in the manifestation of Fock-space NHSE. The derivation of the effective Hamiltonian of the high-energy sector can be found in Appendix \ref{app:N2}.
 
For the half-filled case with $N=L/2$, HSF is significantly enriched by the diverse occupation configurations of particles, while real-space NHSE is suppressed due to the fermionic statistics. We find that the lowest-energy sector forms a HN model in the Fock space, displaying Fock-space NHSE, and the highest-energy sector contains eigenstates that are frozen at single Fock states. On the other hand, the intermediate-energy sectors are far more sophisticated, fragmented into more subsectors with nearly degenerate eigenenergies. The abstract Fock-space lattices for these subsectors generally do not process a uniform non-reciprocity,  yet skin-like localization still persists, with eigenstates localized at the abstract lattice sites associated with the longest hopping sequences exhibiting unidirectional non-reciprocity.  We note that these features are demonstrated by an example with $L=6$ in the main text, while further results and discussion for larger $L$ can be found in Appendix \ref{sec:large_L}.

\section{Acknowledgements}
We thank Yi Qin for helpful discussions. This work is supported by the National Natural Science Foundation of China (Grant No. 12474159), the Fundamental Research Funds for the Central University, Sun Yat-sen University (Grant No. 24qnpy119), and the China Postdoctoral Science Foundation (Grant No. 2024T171067).

\appendix

\section{Perturbation Theory in the High-Energy Sector for $N=2$}\label{app:N2}
\subsection{First-order pertuabtion}\label{app:N2_1st}
In this section, we apply perturbation theory to explain why, under strong interaction $V$, the particle density distribution in the high-energy sector is only sensitive to the variations in $\gamma_2$ and is insensitive to the variations in $\gamma_1$. 
The non-Hermitian Hamiltonian we consider, which includes both NN and NNN hopping, is given by:
\begin{align}
    \hat{H} &= \sum_{x=1}^{L-1} [ (J_1 - \gamma_1)\hat{c}_x^{\dagger}\hat{c}_{x+1} + (J_1 + \gamma_1)\hat{c}_{x+1}^{\dagger}\hat{c}_{x}] \cr
    &+ \sum_{x=1}^{L-2} [(J_2 - \gamma_2)\hat{c}_x^{\dagger}\hat{c}_{x+2} + (J_2 + \gamma_2)\hat{c}_{x+2}^{\dagger}\hat{c}_{x}] \cr
   &+ V\sum_{x=1}^{L-1}\hat{n}_{x}\hat{n}_{x+1}.\label{eq:H_sup}
\end{align}
We treat the interaction term
\begin{align}
\hat{H}_{\rm int} = V\sum_{x=1}^{L-1}\hat{n}_{x}\hat{n}_{x+1}
\end{align}
as the non-perturbative Hamiltonian, and the hopping term
\begin{align}
\hat{H}_{\rm hop} &= \sum_{i=1}^{L-1} [ (J_1 - \gamma_1)\hat{c}_x^{\dagger}\hat{c}_{x+1} + (J_1 + \gamma_1)\hat{c}_{x+1}^{\dagger}\hat{c}_{x}] \cr
    &+ \sum_{x=1}^{L-2} [(J_2 - \gamma_2)\hat{c}_x^{\dagger}\hat{c}_{x+2} + (J_2 + \gamma_2)\hat{c}_{x+2}^{\dagger}\hat{c}_{x}] \cr
\end{align}
as perturbation. 
The unperturbed eigenstates are thus given by two-particle Fock states, which are degenerate either at $E=0$ or $E=V$.
The later constitute the high-energy sector, with basis of this subspace defined as
\begin{align}
    \ket{\alpha_i} = \hat{c}_x^{\dagger} \hat{c}_{x+1}^{\dagger} \ket{0}, \quad x = 1,2,...,L-1.
\end{align}

We can then define a projection operator $\hat{P}_{\rm int} = \sum_{x=1}^{L-1}\ket{\alpha_x}\bra{\alpha_x}$ to project the Hamiltonian onto the high-energy subspace. Under strong interaction $V$, the effective Hamiltonian can be expressed as:
\begin{align}
    &\hat{H}^{(1)}_{\rm eff} = \sum_{x,x' = 1}^{L-1}\ket{\alpha_{x'}}\bra{\alpha_x}\bra{\alpha_{x'}}\hat{H}\ket{\alpha_x}  + \mathcal{O}(\frac{1}{V})\cr
    &= V + \sum_{x,x' = 1}^{L-1}\ket{\alpha_{x'}}\bra{\alpha_x}\bra{\alpha_{x'}}\hat{H}_{\rm hop}\ket{\alpha_x} + \mathcal{O}(\frac{1}{V}).\label{Heff}
\end{align}
It is evident that the contribution from the NN hopping term must be zero, because it will cause $\ket{\alpha_x}$ to jump out of the high-energy sector, resulting in a zero inner product with $\bra{\alpha_{x'}}$. In contrast, the contribution from the NNN hopping term is clearly nonzero, as certain NNN hopping actions will transform $\ket{\alpha_x}$ into $\ket{\alpha_{x \pm 1}}$, leading to a nonzero inner product with $\bra{\alpha_x'}$. Specifically:
\begin{align}
    &\hat{H}_{\rm hop}\ket{\alpha_x} \cr
    &\cong \sum_{x''=1}^{L-2} \left[(J_2 - \gamma_2)\hat{c}_{x''}^{\dagger}\hat{c}_{x''+2} + (J_2 + \gamma_2)\hat{c}_{x''+2}^{\dagger}\hat{c}_{x''}\right] \hat{c}_x^{\dagger}\hat{c}_{x+1}^{\dagger}\ket{0}\cr
    & = \left[(J_2 - \gamma_2)(\hat{c}_{x-2}^{\dagger}\hat{c}_{x+1}^{\dagger}-\hat{c}_{x-1}^{\dagger}\hat{c}_{x}^{\dagger})\right.\cr
    &\left.+ (J_2 + \gamma_2)(\hat{c}_{x+2}^{\dagger}\hat{c}_{x+1}^{\dagger} - \hat{c}_{x+3}^{\dagger}\hat{c}_{x}^{\dagger})\right]\ket{0}\label{1}
\end{align}
Finally, we obtain the matrix elements of the hopping term in the effective Hamiltonian:
\begin{align}
    &\bra{\alpha_{x'}}\hat{H}_{\rm hop}\ket{\alpha_x}\cr
    &= \bra{0}\hat{c}_{{x'}+1}\hat{c}_{x'}\left((J_2 - \gamma_2)(\hat{c}_{x-2}^{\dagger}\hat{c}_{x+1}^{\dagger}-\hat{c}_{x-1}^{\dagger}\hat{c}_{x}^{\dagger})\right.\cr
    &\left.+ (J_2 + \gamma_2)(\hat{c}_{x+2}^{\dagger}\hat{c}_{x+1}^{\dagger} - \hat{c}_{x+3}^{\dagger}\hat{c}_{x}^{\dagger})\right)\ket{0}\cr
    &= -(J_2 - \gamma_2)\delta_{{x'},x-1}
    - (J_2 +\gamma_2)\delta_{{x'},x+1}.
\end{align}
Substituting the above results into \ref{Heff}, the expression for the effective Hamiltonian is:
\begin{align}
    \hat{H}^{(1)}_{\rm eff} &= V - (J_2 - \gamma_2)\ket{\alpha_{x-1}}\bra{\alpha_{x}}\cr
    &- (J_2 + \gamma_2)\ket{\alpha_{x+1}}\bra{\alpha_{x}} + \mathcal{O}(\frac{1}{V}),\label{eq:app_order1}
\end{align}
equivalent to the HN model with NN hopping given by $-(J_2\pm\gamma_2)$. When the interaction $V$ is sufficiently large, higher-order terms can be neglected, so the direction of NHSE is determined solely by the value of $\gamma_2$ and is not sensitive to the variation in $\gamma_1$.

\subsection{Second-order perturbation of NN hopping}\label{app:2nd_per}
As discussed above, NN hopping does not act on the high-energy subspace directly, yet it may still affect it through some higher-order process.
To see this, we consider the second-order perturbation of the model,
\begin{align}
    &\hat{H}^{(2)}_{\rm eff} = V + \hat{P}_{\rm int}\hat{H}_{\rm hop}\hat{P}_{\rm int}  \cr
    &+\hat{P}_{\rm int}\hat{H}_{\rm hop}\frac{1}{V - \hat{H}_{\rm int}}\hat{H}_{\rm hop}\hat{P}_{\rm int} + \mathcal{O}(\frac{1}{V^2}).\label{2}
\end{align}
For the sake of simplicity, we focus on the effect of NN hopping alone, and set $J_2=\gamma_2=0$.
Based on the analysis in the previous section, the first-order perturbation from NN hopping is zero.
The contribution from the second-order perturbation can be obtained as follow:
\begin{align}
    \hat{H}_{\rm hop}\ket{\alpha_x} = \left[ (J_1 - \gamma_1)\hat{c}^{\dagger}_{x-1}\hat{c}^{\dagger}_{x+1} -
    (J_1 + \gamma_1)\hat{c}^{\dagger}_{x+2}\hat{c}^{\dagger}_{x}   \right]\ket{0};
\end{align}

\begin{align}
    &\frac{1}{V - \hat{H}_{\rm int}}\hat{H}_{\rm hop}\ket{\alpha_x} =\cr
    &\frac{1}{V}\left[ (J_1 - \gamma_1)\hat{c}^{\dagger}_{x-1}\hat{c}^{\dagger}_{x+1} -
    (J_1 + \gamma_1)\hat{c}^{\dagger}_{x+2}\hat{c}^{\dagger}_{x}   \right]\ket{0};\cr
\end{align}

\begin{align}
    &\hat{H}_{\rm hop}\frac{1}{V - \hat{H}_{\rm int}}\hat{H}_{\rm hop}\ket{\alpha_x} \cr
    &= \left[\sum_{x=1}^{L-1} (J_1 - \gamma_1)\hat{c}_x^{\dagger}\hat{c}_{x+1} +(J_1 + \gamma_1)\hat{c}_{x+1}^{\dagger}\hat{c}_{x}\right] \cr
    &\times \frac{1}{V}\left[ (J_1 - \gamma_1)\hat{c}^{\dagger}_{x-1}\hat{c}^{\dagger}_{x+1} -
    (J_1 + \gamma_1)\hat{c}^{\dagger}_{x+2}\hat{c}^{\dagger}_{x}   \right]\ket{0}]\cr
    &=\frac{1}{V}\left[(J_1-\gamma_1)^2(\hat{c}^{\dagger}_{x-2}\hat{c}^{\dagger}_{x+1}-\hat{c}^{\dagger}_{x}\hat{c}^{\dagger}_{x-1}) \right. \cr
    &\left.+ (J_1+\gamma_1)^2(\hat{c}^{\dagger}_{x+1}\hat{c}^{\dagger}_{x+2}-\hat{c}^{\dagger}_{x+3}\hat{c}^{\dagger}_{x}) \right.\cr
    &\left.+(J_1^2-\gamma_1^2)(2\hat{c}^{\dagger}_{x}\hat{c}^{\dagger}_{x+1} + 2\hat{c}^{\dagger}_{x-1}\hat{c}^{\dagger}_{x+2})\right]\ket{0};
\end{align}

\begin{align}
    &\bra{\alpha_{x'}}\hat{H}_{\rm hop}\frac{1}{V - \hat{H}_{\rm int}}\hat{H}_{\rm hop}\ket{\alpha_x} = \cr
    &\frac{1}{V}\bra{\alpha_{x'}}\left[(J_1-\gamma_1)^2(\hat{c}^{\dagger}_{x-2}\hat{c}^{\dagger}_{x+1}-\hat{c}^{\dagger}_{x}\hat{c}^{\dagger}_{x-1}) \right.\cr
    &\left.+ (J_1+\gamma_1)^2(\hat{c}^{\dagger}_{x+1}\hat{c}^{\dagger}_{x+2}-\hat{c}^{\dagger}_{x+3}\hat{c}^{\dagger}_{x}) \right.\cr
    &\left.+(J_1^2-\gamma_1^2)(2\hat{c}^{\dagger}_{x}\hat{c}^{\dagger}_{x+1} + 2\hat{c}^{\dagger}_{x-1}\hat{c}^{\dagger}_{x+2})\right]\ket{0}\cr
    &= \frac{1}{V}\left[(J_1 - \gamma_1)^2\delta_{x-1,x'} +(J_1+\gamma_1)^2\delta_{x+1,x'}\right.\cr
    &\left.+ 2(J_1^2-\gamma_1^2)\delta_{x,x'}\right].
\end{align}
Substituting the results into Equation \ref{2} gives:
\begin{align}
    &\hat{P}_{\rm int}\hat{H}_{\rm hop}\frac{1}{V - \hat{H}_{\rm int}}\hat{H}_{\rm hop}\hat{P}_{\rm int} = \frac{1}{V}\bigg(\sum_{x}(J_1+\gamma_1)^2\ket{\alpha_{x+1}}\bra{\alpha_i}\cr
    &+ (J_1 - \gamma_1)^2\ket{\alpha_{x-1}}\bra{\alpha_x} + 2(J_1^2 - \gamma_1^2) \ket{\alpha_x}\bra{\alpha_x}\bigg).
\end{align}
Thus, the expression for $H^{(2)}_{\rm eff}$ is:
\begin{align}
    &H^{(2)}_{\rm eff} =  \frac{1}{V}\sum_{x}\bigg(2(J_1^2 - \gamma_1^2) \ket{\alpha_x}\bra{\alpha_x}+(J_1+\gamma_1)^2\ket{\alpha_{x+1}}\bra{\alpha_x} \cr
    & +(J_1 - \gamma_1)^2\ket{\alpha_{x-1}}\bra{\alpha_x} \bigg)+ V+\mathcal{O}(\frac{1}{V^2}).
\end{align}
The first term describes the second-order process where one of the particles hops to a neighbor site and then hops back,
and the second and third terms describe another second-order process where the two particles hop toward the same direction one after another.
We can see that, compared with Eq.~\eqref{eq:app_order1}, NN hopping affects the high-energy subspace only through a higher-order process of $1/V$, and can therefore be neglected when $V$ is large.

\begin{figure}
    \centering
    \includegraphics[width=1.0\linewidth]{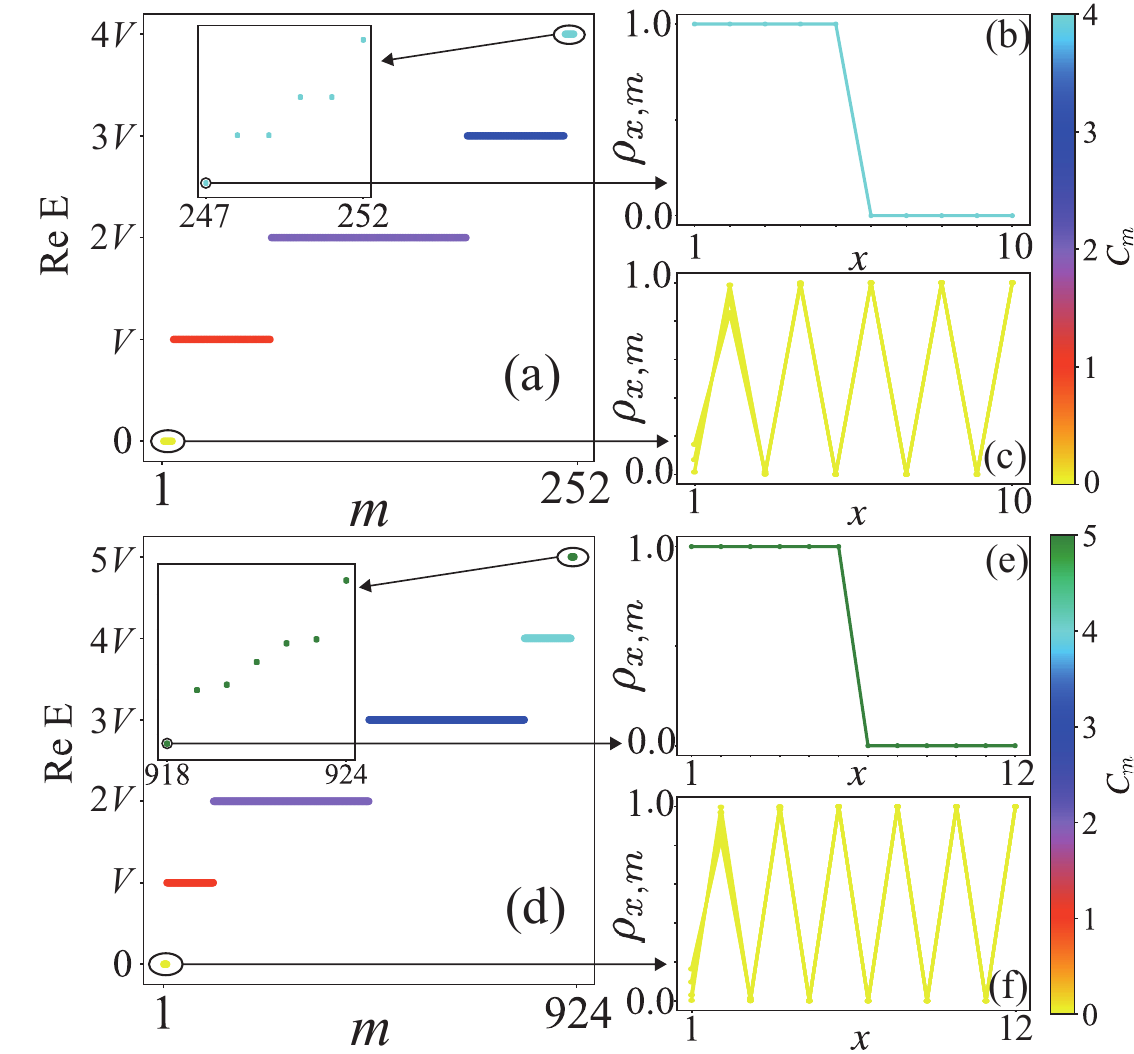}
    \caption{(a) Real parts of eigenenergies for the system $L=10$. Inset shows the zoomed-in view of the highest-energy sector. 
    (b) Particle density distribution of an eigenstate  in the highest-energy sector.
    (c) Particle density distributions of eigenstates in the lowest-energy sector.
    (d) to (e) the same as in (a) to (c), but with $L=12$.
    Parameters are $J_1 =  1.0, \gamma_1 = 0.9, J_2=\gamma_2 = 0, V = 10^{10}$, i.e., the same as in Fig. \ref{fig:half_NN}.
    }\label{fig:half_largeN}
\end{figure}
\section{Half-filling with $L>6$}\label{sec:large_L}
In the main text, we have only considered an example with $L=6$ lattice sites for our model at half-filling.
As the system size increases (with particle number fixed at $L/2$), the Hilbert space is fragmented into more energy sectors by a strong interaction, with real energies ${\rm Re}E\approx 0,V,2V,...,(L/2-1)V$.
Among them, the lowest-energy sector has no adjacent particles, and the highest one has all particles arranged next to each other, similar to Fig. \ref{fig:half_NN}(b) and (d) for $L=6$.
Consequently, eigenstates in the lowest-energy sector also show a density-wave distribution, 
and those in the highest one are frozen at Fock states with particles uniformly distributed on $N$ adjacent lattice sites, 
as illustrated by the examples with $L=10$ and $12$ in Fig. \ref{fig:half_largeN}.

In contrast, the intermedia-energy sectors between them are far more complicated with larger $L$, and the explicit density distribution within each sector needs to be studied case by case. Due to the complicated interplay between NN hopping, NNN hopping, and fermionic statistics,  there may not be a clear pattern of how eigenstates distribute within each intermedia-energy sector, as shown by the example with $L=6$ in the main text.
A common feature of the intermediate-energy sectors is that each sector is further fragmented into a number of disconnected subsectors.
For the case with $L=6$ discussed in the main text, in the absence of NNN hopping, the intermediate-energy sector is fragmented into three subsectors with the same real energies ($E\approx V$).

\begin{figure}[h]
    \centering
    \includegraphics[width=1.0\linewidth]{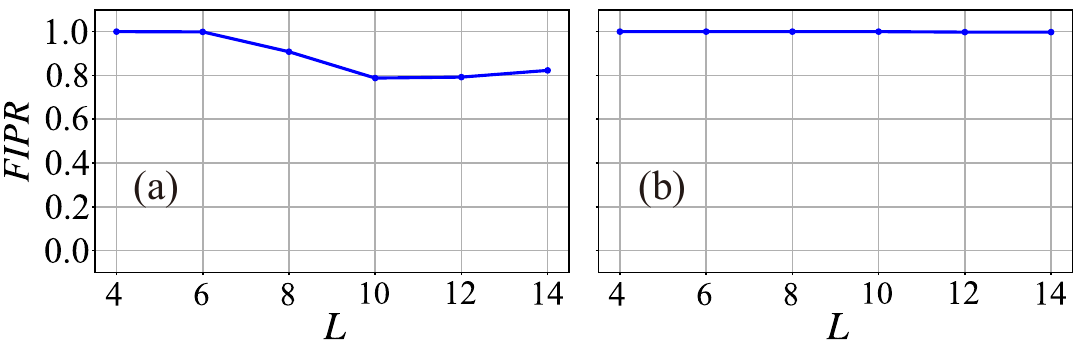}
    \caption{(a) ${\rm FIPR}$ at half-filling as a function of the system's size $L$, in the presence of only NN hopping.
    It drops from one when $L\geqslant 8$, as the number of nearly-degenerate subsectors increases rapidly with $L$.
    (b) ${\rm FIPR}$ with both NN and NNN hopping. It remains to be approximately one, because NNN hopping provide additional connections between different Fock states, and the number of subsectors is therefore much smaller than the dimension of Hilbert space.
    Parameters are the same as in Fig. \ref{fig:half_largeN}, except $J_2=1$ and $\gamma_2=-0.9$ in (b).
    }\label{fig:FIPR}
\end{figure}

However, given that subsectors of the same energy sector are nearly degenerate in energy, they may be sensitive to perturbation such as the higher-order hopping process discussed in Appendix \ref{app:2nd_per}.
A prefect HSF occurs only when $V\rightarrow \infty$, where all higher-order process of $1/V$ vanishes.
Therefore, the many-body eigenstates may not experience a prefect fragmentation of these subsectors [e.g., as the one in Fig.~\ref{fig:half_NN}(f) and (g) for $L=6$], even under a strong but not infinite interaction.
To characterize the degree of HSF,
we define the inverse participation ratio regarding the HSF as
\begin{align}
   &{ {\rm FIPR} }= \frac{1}{M}\sum_{m=1}^{M}{\rm FIPR}_m \cr
  &= \frac{1}{M}\sum_{m=1}^{M}\sum_{s=1}^{S}\left(\sum_{\beta_s=1}^{N_s}|\left\langle \psi_{m} \middle| \beta_s\right\rangle|^2\right)^2,
\end{align}
which describes the degree of ``localization" within each fragment of the Hilbert space, averaged over all eigenstates (labeled by $m$).
Here $S$ is the total number of subsectors,  
$N_s$ is the dimension of Fock basis in subsector $s$, $M$ is the total number of eigenstates.
A prefect HSF is characterized by ${\rm FIPR}=1$, meaning that each eigenstate only occupies Fock states of a single subsector.
As can be seen in Fig. \ref{fig:FIPR}, under a large but finite interaction $V=10^{10}$, ${\rm FIPR}$ drops below one when $L\geqslant8$ for the case with only NN hopping, as the number of nearly-degenerate subsectors increase rapidly with $L$.
On the other hand, in the presence of NNN hopping, ${\rm FIPR}$ remains to be $1$ for $L$ up to $14$, as most nearly-degenerate subsectors are now connected into several non-degenerate sectors by the NNN hopping.

Note that when $L\geqslant 12$, nearly-degenerate subsectors also emerge in the presence of both NN and NNN hopping.
For example, when $L=12$, the intermedia-energy sector with $E\approx 4V$ contains Fock states with two sets of three adjacent particles, separated from each other by three empty lattice sites. Unless longer-range hopping is taken into account, these Fock states are frozen and disconnected from each other and all other states, therefore each of them forms an isolated subsector.
However, their number is much less than the dimension of the complete Hilbert space ($10$ frozen Fock states in a  $C_{12}^{6}$-dimension Hilbert space for $L=12$).
Thus they do not affect the FIPR much even when their HSF is destroyed by perturbation.

\bibliography{ref}

\end{document}